\documentclass[12pt]{article} 
\pdfoutput=1
\usepackage[paper=a4paper, tmargin=20mm, bmargin=20mm, inner=20mm, outer=20mm]{geometry}
\usepackage{graphicx,floatflt,amssymb,rotate}
\usepackage{mathrsfs}
\usepackage{amssymb}
\usepackage{amsmath}
\usepackage{color}
\usepackage{graphicx}
\usepackage{subfig}
\usepackage{multirow}
\usepackage{nonfloat}
\usepackage{longtable}
\usepackage{footnote}
\usepackage{pstricks}
\usepackage[toc,page]{appendix}
\usepackage[mathscr]{euscript}
\usepackage{cite}
\usepackage[colorlinks=true,
linkcolor=red,
urlcolor=blue,
citecolor=blue]{hyperref}
\usepackage{color}

\def\mET{E_T \hspace{-1.0em}/\;\:}

\def\beq{\begin{equation}}
\def\eeq{\end{equation}}
\def\bea{\begin{eqnarray}}
\def\eea{\end{eqnarray}}
\def\nn{\nonumber}

\def\roughly#1{\mathrel{\raise.3ex\hbox
{$#1$\kern-.75em\lower1ex\hbox{$\sim$}}}}

\def\sla#1{\raise.15ex\hbox{$/$}\kern-.57em #1}% Feynman slash

\def\bd{B_d^0}

\def\order{\lower 1.8ex \hbox{\LARGE\~{}}}

\newcommand{\ba}{\begin{array}}
\newcommand{\ea}{\end{array}}

\newcommand{\mptvec}{ \, \not \! \vec{P}_T}

\def\bd0tau{B\to D \tau\nu_{\tau}}

\def\be {\begin{equation}}
\def\ee {\end{equation}}

 \usepackage[normalem]{ulem}

 \definecolor{darkgreen}{cmyk}{1,0,1,0.4}
 
 \definecolor{pink}{cmyk}{0.4,1,0.3,0}

\def\com2#1{\textcolor{red}{\it{#1}}}
\begin{document}

\begin{center}
	
	%opening
	{\Large \bf Collider signature of $V_2$ Leptoquark with $b\to s$  flavour observables}\\
	
\vspace*{1cm}  
\renewcommand{\thefootnote}{\fnsymbol{footnote}}  
{ {\sf Aritra Biswas$^1$ \footnote{email: tpab2@iacs.res.in}},  
{\sf Avirup Shaw$^2$ \footnote{email: avirup.cu@gmail.com}}
and 
{\sf Abhaya Kumar Swain$^1$ \footnote{email: abhayakumarswain53@gmail.com}}
}\\  
\vspace{10pt}  
{ {\em $^1$School of Physical Sciences, Indian Association for the Cultivation of Science,\\
		2A $\&$ 2B Raja S.C. Mullick Road, Jadavpur, Kolkata 700 032, India}\\

\vspace{5pt}  
{\em $^2$ Theoretical Physics, Physical Research Laboratory,\\ 
 Ahmedabad 380009, India}
}
\normalsize  
 
\end{center}

\begin{abstract}
The Leptoquark model has been instrumental in explaining the observed lepton flavour universality violating charged ($b\to c$) and neutral ($b\to s$) current anomalies that have been the cause for substantial excitement in particle physics recently. In this article we have studied the role of one (designated as $V_2^{\frac 43}$) of the components of {\boldmath${V}_2$} Vector Leptoquark doublet with electromagnetic charge $\frac 43$ in explaining the neutral current ($b\to s$) anomalies $R_{K^{(*)}}$ and $B_s\to\mu^+\mu^-$. Moreover, we have performed a thorough collider search for this $V_2^{\frac 43}$ Leptoquark using $b\bar{b} \ell^+ \ell^-$ ($\ell\equiv e, \mu$) final state at the Large Hadron Collider. From our collider analysis we maximally exclude the mass of the $V_2^{\frac 43}$ Leptoquark up to 2340 GeV at 95\% confidence level for the 13 TeV Large Hadron Collider for an integrated luminosity of 3000 ${\rm fb}^{-1}$. Furthermore, a significant portion of the allowed parameter space that is consistent with the neutral current ($b\to s$) observables is excluded by collider analysis.
\end{abstract}

%\maketitle

%\begin{keyword}
%LFUV\sep Leptoquark\sep Collider\sep LHC%\sep sample
%\doi{04.2019/LHEP000001}
%\end{keyword}

%\renewcommand{\thesection}{\Roman{section}}  
\setcounter{footnote}{0} 
\renewcommand{\thefootnote}{\arabic{footnote}}
%\renewcommand{\theequation}{\arabic{equation}} 
%\renewcommand{\theequation}{\arabic{section}.\arabic{equation}}
%%%%%%%%%%%%%%%%%%%%%%%%%%%%%%%%%%%%%%%%%%%%%
\section{Introduction}\label{intro}
%%%%%%%%%%%%%%%%%%%%%%%%%%%%%%%%%%%%%%%%%%%%%

The discovery of the Higgs boson in 2012 by the CMS~\cite{Chatrchyan:2012xdj} and ATLAS~\cite{Aad:2012tfa} collaborations is definitely one of the greatest achievements of Large Hadron Collider (LHC). Unfortunately, it has not been able to detect signatures corresponding to any new physics (NP) particles till date. On the other hand experimental measurements of observables related to $B$ physics have, however, exhibited deviations of a few $\sigma$ from their Standard Model (SM) expectations hinting towards the existence\footnote{Apart from such deviations, non-zero neutrino mass, signatures for the existence of dark matter, observed baryon asymmetry etc. also concur to the fact that BSM physics is indeed a reality of nature.} of beyond SM (BSM) physics. $B$-physics experiments at LHCb, Belle and Babar have pointed at intriguing lepton flavour universality violating (LFUV) effects. To that end, flavour changing neutral current\footnote{Experimental signatures are also present for LFUV via charge current semileptonic $b \to c \ell \nu$ transition processes. For example the ratios $R_{D^{(*)}}$~\cite{average} and $R_{J/\psi}$~\cite{Aaij:2017tyk} show significant deviations from their corresponding SM predictions.} (FCNC) processes such as $b \to s \mu^+ \mu^-$ have drawn much attention due to anomalies that have been observed recently at the LHCb and Belle experiments. A deviation of 2.6$\sigma$ has been observed in $R_K = {\rm BR}(B^+\to K^+ \mu^+ \mu^-)/{\rm BR}(B^+\to K^+ e^+ e^-)$ with a value of $0.745^{+0.090}_{-0.074}\pm 0.036$~\cite{Aaij:2014ora} from the corresponding SM prediction ($R_K\approx 1$~\cite{Descotes-Genon:2015uva,Bordone:2016gaq}) for the integrated di-lepton invariant mass squared range $1 < q^2 < 6$ ${\rm GeV}^2$. LHCb has reported a deviation in $R_{K^*} = {\rm BR}(B^0 \to K^{*0} \mu^+ \mu^-)/{\rm BR}(B^0 \to K^{*0} e^+ e^-)$ at the level of  $2.1-2.3\sigma$ and $2.4-2.5\sigma$ for the two $q^2$ ranges [0.045-1.1] (called low-bin) and [1.1-6.0] ${\rm GeV^2}$ (called central-bin) with values $0.660^{+0.110}_{-0.070} \pm 0.024$~\cite{Aaij:2017vbb} and  $0.685^{+0.113}_{-0.069} \pm 0.047$~\cite{Aaij:2017vbb} respectively. The corresponding SM predictions are $0.92\pm 0.02$~\cite{Capdevila:2017bsm} and $1.00\pm 0.01$~\cite{Descotes-Genon:2015uva,Bordone:2016gaq} respectively.

In order to explain the above mentioned anomalies we have selected a particular extension of the SM consisting of several hypothetical particles that mediate interactions between quarks and leptons at tree-level. Hence, these particles are known as Leptoquarks (LQs). Such particles can appear naturally in several extensions of the SM~(e.g., composite models \cite{Schrempp:1984nj}, Grand Unified Theories \cite{Georgi:1974sy, Pati:1973rp, Dimopoulos:1979es, Dimopoulos:1979sp, Langacker:1980js, Senjanovic:1982ex, Cashmore:1985xn, Pati:1974yy}, superstring-inspired ${\rm E_6}$ models~\cite{Green:1984sg, Witten:1985xc, Gross:1984dd, Hewett:1988xc} etc). Considerable amount of work regarding LQs have been done both from
the point of view of their diverse phenomenological aspects~\cite{Davidson:1993qk,Hewett:1997ce,Nath:2006ut}, 
and specific properties~\cite{Shanker:1981mj,Shanker:1982nd,Buchmuller:1986iq,
Buchmuller:1986zs,Hewett:1987yg,Leurer:1993em,Leurer:1993qx, Aaltonen:2007rb,
Dorsner:2014axa,Allanach:2015ria,Evans:2015ita,Li:2016vvp, Diaz:2017lit,
Dumont:2016xpj,Faroughy:2016osc,Greljo:2017vvb, Baumgartel:2014dqa,
Aad:2015caa,Aaboud:2016qeg,Sirunyan:2017yrk,Sirunyan:2018nkj, 
Dorsner:2017ufx,Allanach:2017bta,Crivellin:2017zlb,
Hiller:2017bzc,Buttazzo:2017ixm,Calibbi:2017qbu,Sahoo:2016pet,
Altmannshofer:2017poe,Biswas:2018snp, Blanke:2018sro}. Furthermore, several articles ~\cite{Sakaki:2013bfa, Becirevic:2016yqi, Popov:2016fzr, Chen:2017hir, Becirevic:2017jtw, Alok:2017sui, Crivellin:2017zlb, Assad:2017iib, Aloni:2017ixa,Wold:2017wdj,Muller:2018nwq,Hiller:2018wbv,Biswas:2018jun,Fajfer:2018bfj,Monteux:2018ufc, Becirevic:2018uab, Kumar:2018kmr, Crivellin:2018yvo, Angelescu:2018tyl} that explain the different flavour anomalies with different versions of LQ models exist in the literature.

In connection to the above, we consider one of the components of the {\boldmath${V}_2$} vector leptoquark (VLQ) doublet (the $V_2^{\frac 43}$) that is capable of mediating the $b\to s$ observables at tree level, due to its electromagnetic charge $Q=\frac43$. We provide bounds on the parameter space for the $V_2^{\frac 43}$ VLQ subject to constraints due to the observables $R_{K^{(*)}}$. Furthermore, we have used the latest experimental value $2.8^{+0.7}_{-0.6}\times 10^{-9}$~\cite{CMS:2014xfa} of the branching fraction for the decay $B_s\to\mu^+\mu^-$ as another constraint in our analysis while the SM prediction for the same decay is $3.66\pm 0.23\times 10^{-9}$~\cite{Bobeth:2013uxa}. Out of the eight Wilson coefficients (given in eq.\;\ref{WC} of sec.\;\ref{flav}) that contribute to the above $b\to s$ observables mediated by the $V_2^{\frac 43}$ VLQ, only four are independent. This allows us to numerically solve for these coefficients and in turn provide constraints on the real and imaginary parts for the allowed values of the coupling products $(g_{L,R})_{b\ell}(g_{L,R})_{s\ell}$ with respect to the mass of the $V_2^{\frac 43}$ VLQ up to $1\sigma$ (corresponding to the $1\sigma$ experimental errors for these observables).

The LQs being potential candidates in explaining the flavour anomalies, it is only relevant that one investigates the production and decay signatures of these LQs at colliders. There exist several articles~\cite{Dorsner:2014axa,Allanach:2015ria,Evans:2015ita,Diaz:2017lit,Greljo:2017vvb, Dorsner:2017ufx, Bandyopadhyay:2018syt, Vignaroli:2018lpq} in the literature that have been dedicated to collider studies of LQs, but in most of the cases these studies have been performed on scalar LQs. The collider studies for vector LQs are limited in number~\cite{Aaltonen:2007rb, Diaz:2017lit, Biswas:2018snp, Dorsner:2018ynv, Sirunyan:2018ruf}. Our current interest for this article being the $V^{\frac 43}_2$ VLQ, it is imperative that one probes this LQ at the current or future collider experiments. To the best of our knowledge, the present article is the first which deals with the collider prospects of the $V^{\frac 43}_2$ VLQ\footnote{The {\boldmath${V}_2$} VLQ belongs to the anti-fundamental representation of the ${\rm SU}(3)_{\rm C}$ part of the SM gauge group~\cite{Dorsner:2018ynv}. Hence, there is no available model file for this VLQ. Therefore, we believe this to be the first article which deals with collider prospects of {\boldmath${V}_2$} VLQ after proper implementation of the model in {\sl FeynRules}~\cite{Alloul:2013bka}.} at the LHC.  We study signatures corresponding to this VLQ for $b\bar{b} \ell^+ \ell^-$ final states at the LHC with the centre of momentum (CM) energy $\sqrt{s}=13$ TeV. Although the ATLAS collaboration has also looked at the same final state~\cite{ATLAS:2017hbw} but they have searched for the R-parity violating scalar top partners at the 13 TeV LHC. Their exclusion limit, depending on the branching fractions of the scalar top to bottom and electron/muon, is set from 600 GeV to 1500 GeV. Using several interesting kinematic variables we maximize the signal event with respect to relevant SM backgrounds. From our collider analysis and depending on the SM bilinear couplings with $V^{\frac 43}_2$ VLQ we exclude the mass of this VLQ up to 2140 GeV and 2340 GeV for the two bench mark values of integrated luminosities 300 ${\rm fb}^{-1}$ and 3000 ${\rm fb}^{-1}$ respectively at the 95\% confidence level (C.L.). At this point, we would like to mention that the other component ($V_2^{\frac 13}$) of the {\boldmath${V}_2$} VLQ with electromagnetic charge $Q=\frac13$ has not been considered in this analysis primarily because it is unable to mediate the $b\to s \ell^+ \ell^-$ interactions. In addition, the parameter values taken in this analysis result in a small value of the branching ratio of $V_2^{\frac 13}$ VLQ to up type quarks and charged leptons or any final state. Hence the collider reach would be weak compared to the signal we have considered.
    
The paper is organised as follows. We briefly discuss the Lagrangian for the {\boldmath$V_2$} VLQ and set the notations in section~\ref{v2model}. In section~\ref{flav} we show the flavour analysis of $b \rightarrow s$ transition observables mediated by the $V^{\frac 43}_2$ VLQ. Section~\ref{collider} is dedicated to the collider analysis for $V^{\frac 43}_2$ with $b\bar{b} \ell^+ \ell^-$ final states. Finally, we discuss our results and conclude in section \ref{con}.

%%%%%%%%%%%%%%%%%%%%%%%%%%%%%%%%%%%%%%%%%%%%%
\section{Effective Lagrangian of {\boldmath${V_2}$} vector Leptoquark}\label{v2model}
%%%%%%%%%%%%%%%%%%%%%%%%%%%%%%%%%%%%%%%%%%%%%
LQs are special kinds of hypothetical particles that carry both lepton (L) and baryon (B) number. Consequently they couple to both leptons and quarks simultaneously. Furthermore, they possess colour charge and fractional electromagnetic charges. However, unlike the quarks they are either scalars or vectors bosons. For further discussions regarding all LQ scenarios, one can  look into the review~\cite{Dorsner:2016wpm}. Due to the above distinguishable properties, these LQs have several phenomenological implications with respect to the other BSM particles. In general there are twelve LQs, among them six are scalars ($S_3$, {$R_2$}, $\tilde{R_2}, \tilde{S_1}, S_1, \bar{S_1}$) and the rest ($U_3$, {\boldmath$V_2$}, $\tilde{V_2}, \tilde{U_1}, U_1, \bar{U_1}$) transform vectorially under Lorentz transformations. In the current article we are particularly interested on {\boldmath${V_2}$} VLQ in order to explain the $b\to s$ anomalies. Under the SM gauge group ${\rm {SU(3)_C\times SU(2)_L\times U(1)_Y}}$ the {\boldmath${V_2}$} VLQ transforms as $({\bf{\bar{3},2}},\frac56)$. The Lagrangian which describes the interaction for the {\boldmath${V_2}$} VLQ with the SM fermion bilinear is given as~\cite{Dorsner:2016wpm}
\begin{equation}
\mathcal{L}^{\rm LQ}_{\boldsymbol{V_2}}= (g_{L})_{ij}\bar{d}_{iR}^C\gamma^\mu {\boldsymbol{V_2}}^a_{,\mu}\epsilon^{ab}L^b_{jL}+(g_{R})_{ij}\bar{Q}_{iL}^{C,a}\gamma^\mu \epsilon^{ab} {\boldsymbol{V_2}}^b_{,\mu}\ell_{jR} + {\rm h.c.},
\label{Lagrangianv2}
\end{equation}
with $a, b\equiv 1, 2$. Here, $Q^{\sf T}_L\equiv (u~~~d)$ represents the left handed quark doublet, $L^{\sf T}_L\equiv (\nu_{\ell}~~~\ell)$ denotes for the left handed lepton doublet, $d_R$ stands for the right handed down type quark singlet and $\ell_R$ is the right handed charged lepton singlet. Left (right) handed gauge coupling constants are represented by $(g_{L(R)})_{ij}$ with the fermion generation indices $i, j\equiv 1,2,3$. To avoid the constraint due to the proton decay from {\boldmath${V_2}$} VLQ, we set the corresponding {\boldmath${V_2}$} couplings for di-quark interactions to zero\footnote{Since we work in an effective framework and not an ultraviolet (UV) complete model in the current article, we can hence treat the couplings as free parameters.}. As the {\boldmath${V_2}$} VLQ is transformed as doublet under ${\rm SU(2)_L}$ gauge group hence this {\boldmath${V_2}$} VLQ multiplet contains two components $V_2^{\frac 43}$ and $V_2^{\frac 13}$ having electromagnetic charges $\frac 43$ and $\frac 13$ respectively. In the following we will focus only on the one component $V_2^{\frac 43}$ carrying electromagnetic charge $\frac 43$. From hereon, we will refer to the $V_2^{\frac 43}$ VLQ simply as $V_2$. 

%%%%%%%%%%%%%%%%%%%%%%%%%%%%%%%%%%%%%%%%%%%%%
\section{Flavour signatures}\label{flav}
%%%%%%%%%%%%%%%%%%%%%%%%%%%%%%%%%%%%%%%%%%%%%
We closely follow ref.~\cite{Kosnik:2012dj} in the following discussion about the operator basis relevant to $b\to s\ell^+\ell^-$ decays and the expressions for the observables. The effective dimension six Hamiltonian at the mass scale of the $b$ quark is written as~\cite{Kosnik:2012dj, Grinstein:1988me}
 \begin{eqnarray}
  \mathcal{H}_{eff}&=&-\frac{4G_F}{\sqrt{2}}\lambda_t\bigg[\sum_{i=1}^6C_iO_i+
\sum_{i=7,8,9,10,P,S}{\hspace*{-0.8cm}}(C_iO_i+C_i^\prime(\mu)O_i^\prime(\mu))+C_TO_T+C_{T5}O_{T5}\bigg],
  \label{ham}
 \end{eqnarray}
where $\lambda_t=V_{tb}V_{ts}^*$. The $V_2$ VLQ contributes to the following two-quark, two-lepton operators:

\begin{eqnarray}
 O_9&=&\frac{e^2}{g^2}(\bar{s}\gamma_\mu P_Lb)(\bar{\ell}\gamma^\mu \ell)\;,~~%\nonumber\\
 O_{10}=\frac{e^2}{g^2}(\bar{s}\gamma_\mu P_Lb)(\bar{\ell}\gamma^\mu\gamma_5 \ell)\;,\nonumber\\
 O_S&=&\frac{e^2}{16\pi^2}(\bar{s}P_Rb)(\bar{\ell}\ell)\;,~~%\nonumber\\
 O_P=\frac{e^2}{16\pi^2}(\bar{s}P_Rb)(\bar{\ell}\gamma_5\ell)\;,
 \label{op}
\end{eqnarray}
and their corresponding ``primed'' counterparts.  The chiraly flipped ``primed'' operators are obtained by an $L\leftrightarrow R$ exchange in the above operators. Here $e=\sqrt{4\pi\alpha}$ represents the unit for electromagnetic charge, $g$ is the strong coupling constant and $P_{R,L}=(1\pm\gamma_5)/2$. The four-quark operators $O_{1-6}$ and the radiative penguin operators $O_{7,8}$ are provided in ref.~\cite{Bobeth:1999mk}. The decay amplitudes for the $B\to K^*\ell^+\ell^-$ transition in terms of the effective Wilson coefficients (WCs) evaluated at the scale $\mu=m_b$ are provided in~\cite{Buras:1993xp}.

The theoretical expression for the branching fraction corresponding to the $B_s\to \ell^+\ell^-$ decay reads~\cite{Kosnik:2012dj}
\begin{eqnarray}
 {\rm BR}(B_s\to \ell^+\ell^-)&=&\tau_{B_s}f_{B_s}^2m_{B_s}^3\frac{G_F^2|\lambda_t|^2\alpha^2}{(4\pi)^3}\beta_\ell(m_{B_s}^2)
 \bigg[\frac{m_{B_s}^2}{m_b^2}|C_S-C_{S^\prime}|^2(1-\frac{4m_\ell^2}{m_{B_s}^2})+\nonumber\\
&&|\frac{m_{B_s}}{m_b}(C_P-C_{P^\prime})+2\frac{m_\ell}{m_{B_s}}(C_{10}-C_{10}^\prime)|^2\bigg].
 \label{br}
\end{eqnarray}
In the above $\beta_\ell=\sqrt{1-4m_\ell^2/q^2}$, $m_{B_s}$, $m_b$ and $m_\ell$ are denoted as the masses of $B_s$ meson, bottom quark ($b$) and charged lepton ($\ell$) respectively. $G_F$ is the Fermi constant, $\tau_{B_s}$ represents the life time while $f_{B_s}$ stands for the decay constant of $B_s$ meson. It is evident from eq.~\ref{br}, that the ${\rm BR}(B_s\to\mu^+\mu^-)$ (considering $\ell\equiv \mu$) is only sensitive to the contributions due to the differences between operators with left and right-handed quark currents, $C_{10}-C_{10}^\prime$, $C_S-C_S^\prime$ and $C_P-C_P^\prime$.

In contrast to the case for $B_s\to\mu^+\mu^-$, the decay width for $B\to K\ell^+\ell^-$ receives contributions from $C_7+C_7^\prime$, $C_9+C_9^\prime$, $C_{10}+C_{10}^\prime$, $C_S+C_S^\prime$ and $C_P+C_P^\prime$. The tensor operators have small contributions in LQ models~\cite{Kosnik:2012dj}. The corresponding decay width reads~\cite{Bobeth:2007dw}
\begin{equation}
 \Gamma(B\to K\ell^+\ell^-)=2(A_\ell+\frac{1}{3}C_\ell),
\end{equation}
where 
\begin{equation}
 A_\ell=\int_{4m_\ell^2}^{(m_B-m_K)^2}a_\ell(q^2)dq^2,\;\;C_\ell=\int_{4m_l^2}^{(m_B-m_K)^2}c_\ell(q^2)dq^2.
\end{equation}
$a_\ell$ and $c_\ell$ are defined as:
\begin{eqnarray}
a_\ell (q^2) &=&  {\cal C}(q^2)\Big[  q^2 \left( \beta_{\ell}^2(q^2) \lvert F_S(q^2)\rvert^2 + \lvert F_P(q^2) \rvert^2 \right)+\frac{\lambda (q^2) }{4} \left( \lvert F_A (q^2)\rvert^2 +  \lvert F_V(q^2) \rvert^2 \right)\nn \\
&&+ 4 m_{\ell}^2 m_B^2 \lvert F_A(q^2) \rvert^2+2m_{\ell}  \left( m_B^2 -m_K^2 +q^2\right) \text{Re}\left( F_P(q^2) F_A^{\ast}(q^2) \right)\Big]  \,,\nn \\ 
c_\ell (q^2) &=&  {\cal C}(q^2)\Big[ - \frac{\lambda (q^2)}{4} \beta_{\ell}^2(q^2) \left( \lvert F_A(q^2) \rvert^2 +  \lvert F_V(q^2) \rvert^2 \right)  \Big]\,,\nn
\end{eqnarray}
where
\begin{eqnarray}\label{F_BKell}
F_V (q^2)  &=&   \left( C_9 + C_9^{\prime  } \right)  f_+ (q^2) + \frac{2 m_b}{m_B +m_K} \left(   C_7  + C_7^{\prime } \right)  f_T (q^2)   \,,  \nonumber\\ 
F_A (q^2)  &=& \left( C_{10} +C_{10}^{\prime  } \right)  f_+ (q^2) \,,   \nonumber\\
F_S (q^2)  &=& \frac{m_B^2 -m_K^2}{2m_b}  \left( C_S + C_S^{\prime} \right) f_0 (q^2)\,,  \nonumber\\
F_P (q^2)  &=&  \frac{m_B^2 -m_K^2}{2m_b} \left( C_P + C_P^{\prime} \right)  f_0 (q^2) - m_{\ell} \left( C_{10}+C_{10}^{\prime } \right) \,\left[  f_+ (q^2) - \frac{m_B^2-m_K^2}{q^2} \left(f_0  (q^2) - f_+ (q^2) \right) \right]  \,.\nn
%F_T (q^2) =&   \frac{2 \sqrt{\lambda (q^2)} \beta_{\ell}(q^2)}{m_B + m_K}  C_T f_T(q^2)\,,\nn\\
% F_{T5} (q^2)  =&  \frac{2 \sqrt{\lambda (q^2)} \beta_{\ell}(q^2)}{m_B + m_K}  C_{T5}f_T(q^2) \,. 
\end{eqnarray}
Here
\begin{eqnarray}
  \mathcal{C}(q^2) &=& \frac{G_F^2 \alpha^2 \lvert \lambda_t
    \rvert^2}{512 \pi^5 m_B^3}  \beta_{\ell}(q^2)   \sqrt{\lambda (q^2)
  } \,,\\
  \lambda (q^2) &=& q^4+ m_B^4 + m_K^4 -2 \left( m_B^2 m_K^2 +m_B^2 q^2 + m_K^2 q^2 \right) \,.\nn
\end{eqnarray}

The functions $F_i$, for $i=V,A,S,P$ are defined as:
\begin{eqnarray}
\langle K(k)| \bar{s} \gamma_{\mu} b |B(p)\rangle &=&  \left[ (p +
  k)_\mu - {m_B^2 - m_K^2 \over q^2} q_\mu \right] f_{+}(q^2)+ {m_B^2-m_K^2 \over q^2} q_{\mu} f_{0}(q^2) \, ,  \\
\langle K(k)| \bar{s}\sigma_{\mu\nu}b |B(p) \rangle &=& i \left( p_\mu k_\nu  - p_\nu k_\mu \right) \frac{2 f_T(q^2)}{m_B + m_K} \, .
\end{eqnarray}
The form factors $f_+$, $f_0$ and $f_T$ have been obtained from ref.~\cite{Altmannshofer:2014rta} where the authors perform a combined fit to the lattice computation~\cite{Bouchard:2013pna} and light cone sum rules (LCSR) predictions at $q^2=0$~\cite{Ball:2004ye,Bartsch:2009qp}, using the parametrization and conventions of~\cite{Bouchard:2013pna}.

WCs corresponding to the operators related to the $V_2$ VLQ (eq.~\ref{op}) that contribute to a $b\to s \ell^+ \ell^-$ transition are~\cite{Kosnik:2012dj}:
\begin{align}
  C_9 &= C_{10} = \frac{-\pi}{\sqrt{2} G_F \lambda_t \alpha}
  \frac{(g_R)_{b\ell} (g_R)^*_{s\ell}}{M^2_{V_2}}\,,\nn\\
  -C_9^\prime &= C_{10}^\prime = \frac{\pi}{\sqrt{2} G_F \lambda_t \alpha}
  \frac{(g_L)_{b\ell} (g_L)^*_{s\ell}}{M^2_{V_2}}\,,\nn\\
  C_P &= C_{S} = \frac{\sqrt{2}\pi}{G_F \lambda_t \alpha}   \frac{(g_R)_{b\ell} (g_L)^*_{s\ell}}{M^2_{V_2}}\,,\nn\\
  -C_P^\prime &= C_{S}^\prime = \frac{\sqrt{2}\pi}{G_F \lambda_t \alpha}   \frac{(g_L)_{b\ell} (g_R)^*_{s\ell}}{M^2_{V_2}}\,.
  \label{WC}
\end{align}

\begin{figure}[ht!]
\centering
\subfloat[]{\label{dc9}\includegraphics[width=0.5\linewidth]{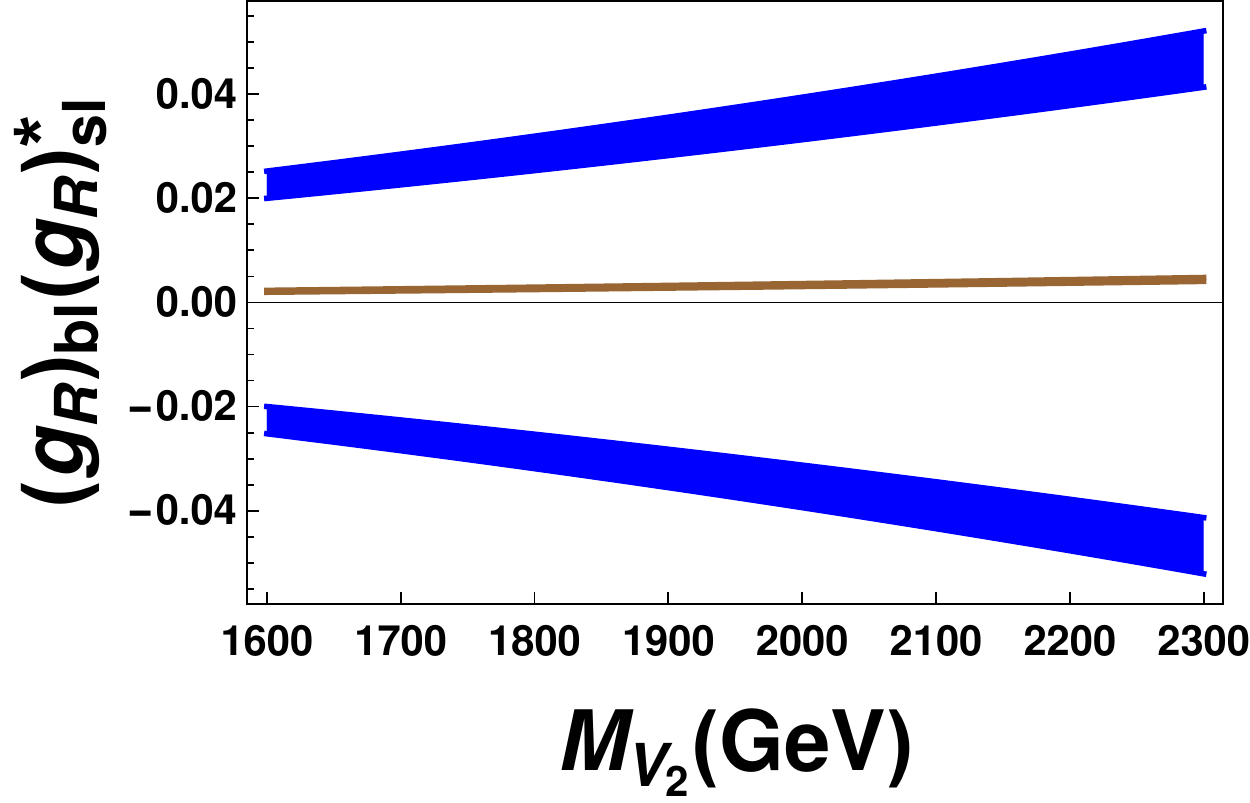}}\hfill
\subfloat[]{\label{c10pr}\includegraphics[width=0.5\linewidth]{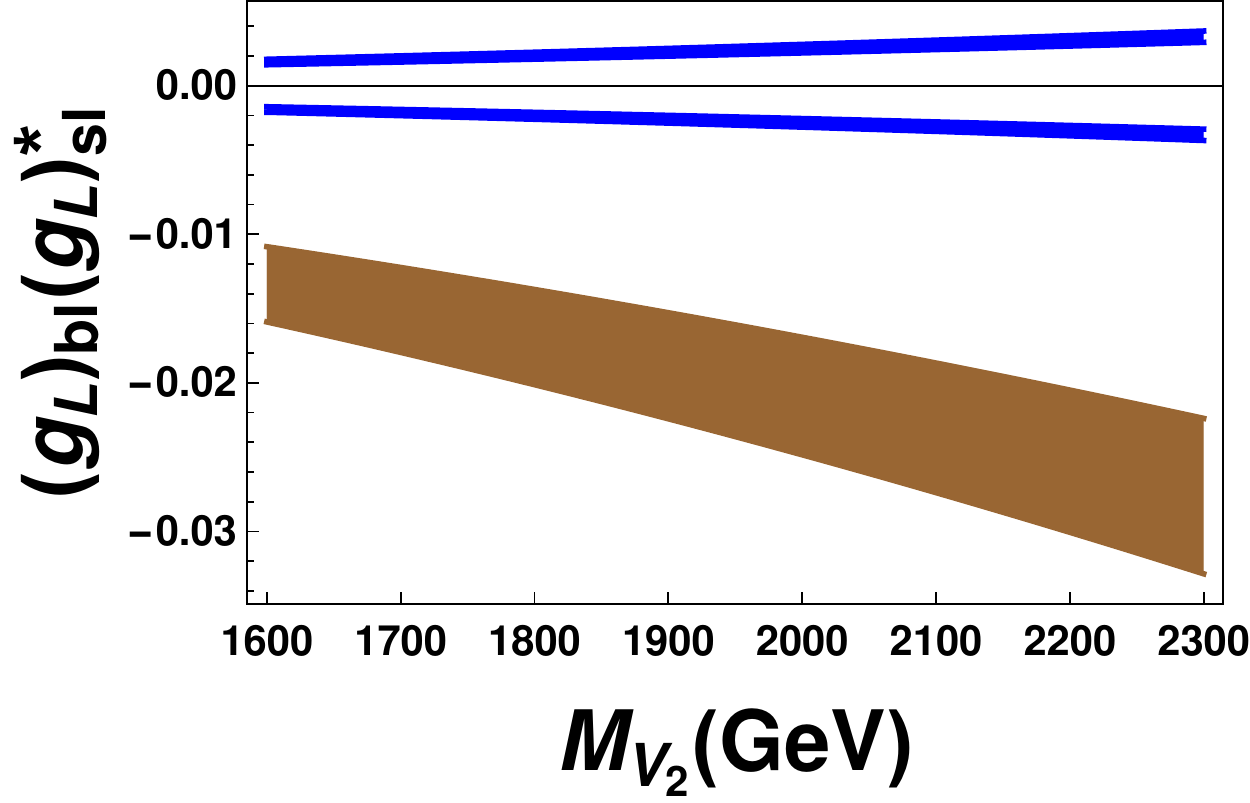}}\\
\subfloat[]{\label{cP}\includegraphics[width=0.5\linewidth]{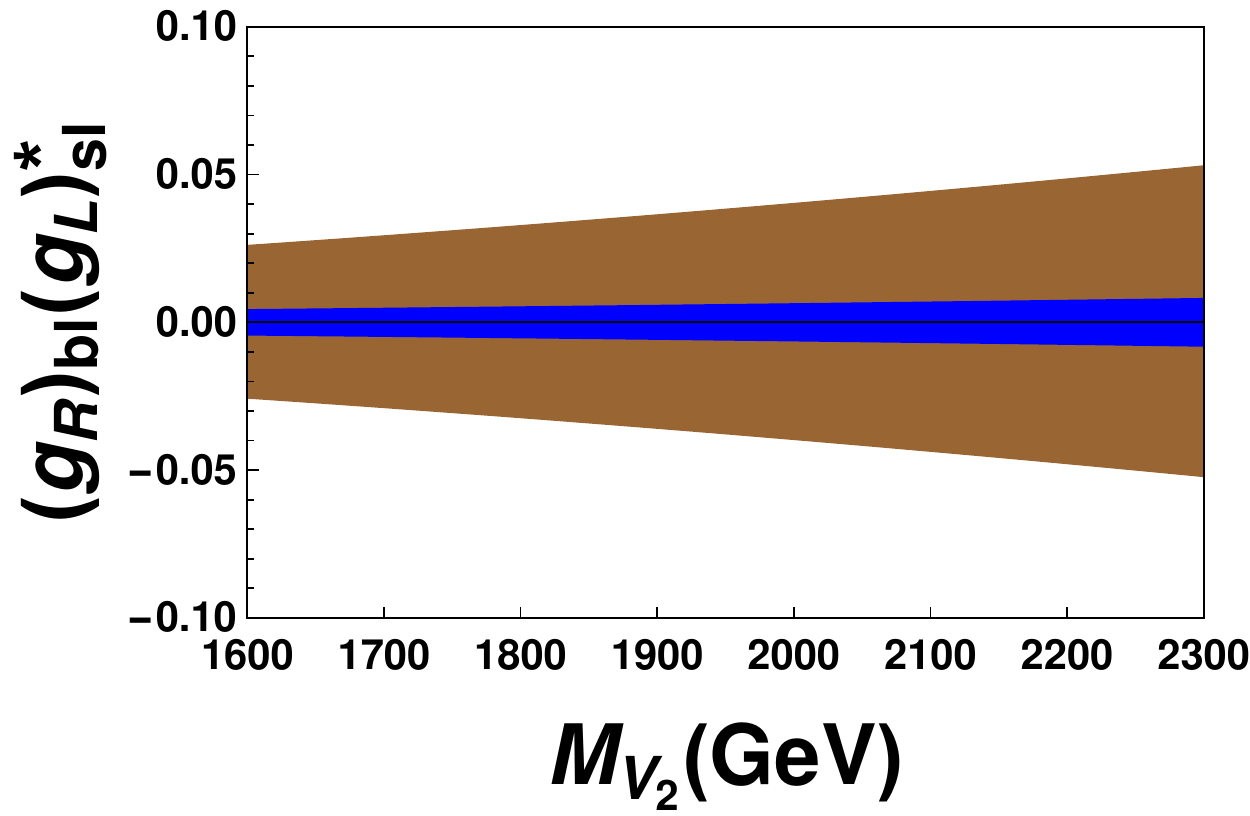}}\hfill
\subfloat[]{\label{cSpr}\includegraphics[width=0.5\linewidth]{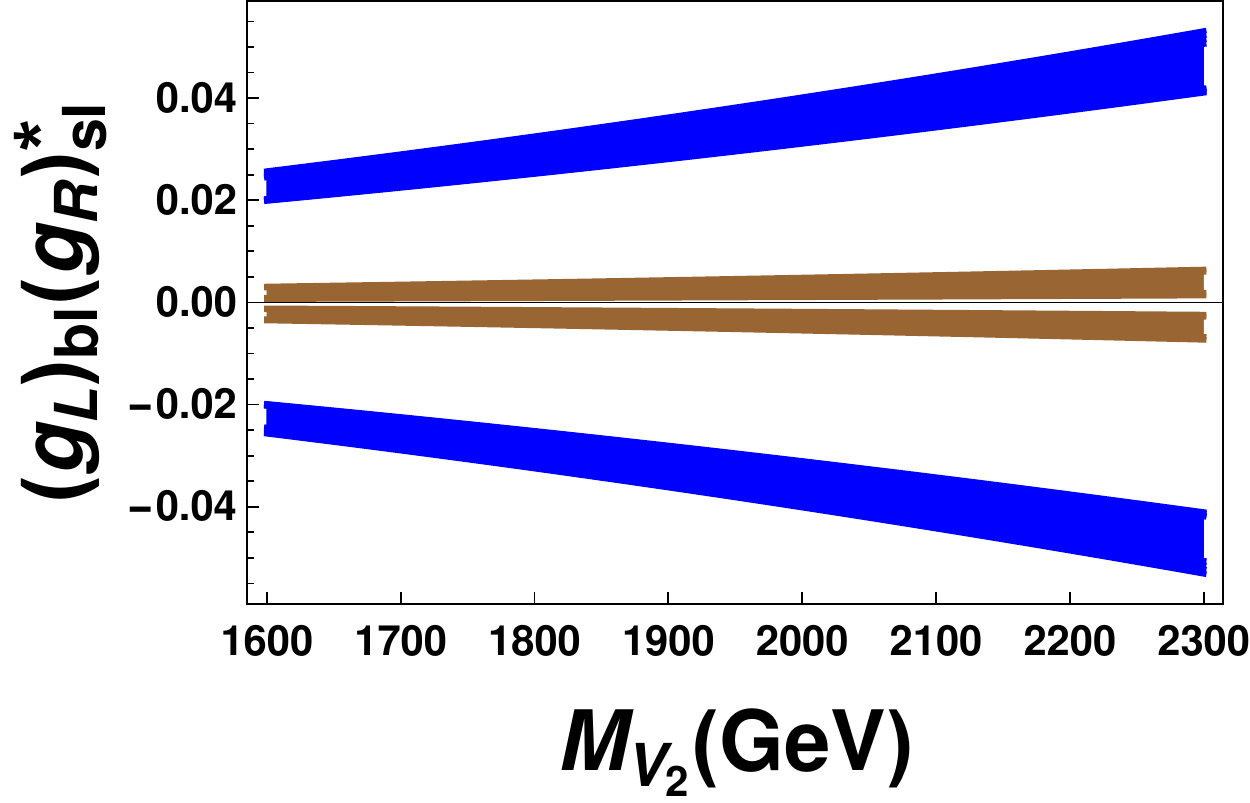}}\\
\subfloat{\includegraphics[width=0.30\linewidth]{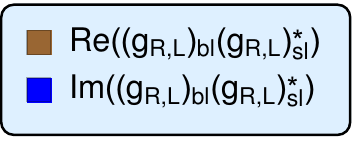}}
\caption{The constraints on the parameter spaces of the $V_2$ VLQ model due to the experimental values provided in introduction. The vertical axes for the plots are products of the model-couplings. In a clockwise fashion, they are: (a)$(g_R)_{b\ell}(g_R)_{s\ell}^*$~, (b)$(g_L)_{b\ell}(g_L)_{s\ell}^*$~, (c)$(g_R)_{b\ell}(g_L)_{s\ell}^*$ and (d)$(g_L)_{b\ell}(g_R)_{s\ell}^*$. The horizontal axis represents the mass of the $V_2$ VLQ in GeV for all four cases. The range for the same is in accordance with the limits obtained from the collider analysis provided in the next section. The brown bands correspond to the real and the blue bands correspond to the imaginary parts of the corresponding coupling products depicted along the vertical axis of each plot. The legend is provided at the bottom. In the above, $\ell=\mu$.} 
\label{flavour_plots}
\end{figure}

It is evident that of the eight relevant WCs, only four are independent, which we take to be $C_9$, $C_{10}^\prime$, $C_P$ and $C_S^\prime$. Although there is a large number of binned data for numerous other observables in the $b\to s$ sector due to LHCb, the four observables that we work with ($R_K$, $R^{\rm low-bin}_{K^{*}}$, $R^{\rm central-bin}_{K^{*}}$ and ${\rm BR}(B_s\to\mu^+ \mu^-)$) are known as the ``clean observables'', i.e. they are precisely measured and suffer from less theoretical uncertainties in comparison to the other observables. Since we have four such observables and four independent WCs, a ``fit'' becomes meaningless and hence we ``solve'' for these coefficients. Hence, these WCs correspond to the values of the observables within their experimental ($1\sigma$) errors exactly. These solutions translate to constraints on the model parameters for the $V_2$ VLQ scenario. These constraints are displayed in fig.~\ref{flavour_plots} for the real and imaginary parts of the coupling product. In general, however, constraints on individual couplings cannot be derived from flavour physics alone since it is the product of the couplings that enter the individual WCs (viz. eq.~\ref{WC}). The bands correspond to the $1\sigma$ experimental errors for the measured observables.

Fig.~\ref{flavour_plots}(a) displays the variation of the real and imaginary parts of the coupling product $(g_R)_{b\ell}(g_R)_{s\ell}^*$ with respect to the mass of the $V_2$ LQ $M_{V_2}$. The variation for the real (imaginary) part is due to the real (imaginary) part of the solution for the WC $C_9$ with respect to the experimental observables given in introduction. The real part of $C_9$ has a unique solution, resulting in the single brown band close to the horizontal axis in fig.~\ref{flavour_plots}(a). However, the imaginary part of $C_9$ has two sets of solutions which are symmetric with respect to 0, and hence translate into the blue bands symmetric with respect to to the horizontal axis. Similarly, the real and imaginary parts for the $C_{10}^\prime$ WC translate into fig.~\ref{flavour_plots}(b). The unique negative solution for the real part translates into the wide brown band and the solutions for the imaginary part give rise to the blue bands symmetric to the horizontal axis for the coupling product $(g_L)_{b\ell}(g_L)_{s\ell}^*$. For a benchmark value $M_{V_2}=1600$ GeV, the ranges for the real and imaginary parts of these coupling products are:
\begin{eqnarray}
{\rm Re}((g_R)_{b\ell}(g_R)_{s\ell}^*)&\in&[0.0019,0.0023],\nn \\
~{\rm Im}((g_R)_{b\ell}(g_R)_{s\ell}^*)&\in& \big([0.020,0.025],
[-0.025,-0.020]\big);\nn \\
{\rm Re}((g_L)_{b\ell}(g_L)_{s\ell}^*)&\in&[-0.016,-0.011],\nn \\
~{\rm Im}((g_L)_{b\ell}(g_L)_{s\ell}^*)&\in&\big([0.0014,0.0018],[-0.0018,-0.0014]\big).\nn 
\end{eqnarray}
The cases~\ref{flavour_plots}(c) and~\ref{flavour_plots}(d) are a little different from the cases discussed above. \ref{flavour_plots}(c) arises due to $C_P$, both of whose real and imaginary part have two solutions, one positive and one negative, at both the higher and lower limits considering experimental errors. However, the regions for these solutions overlap, and hence get broad brown and blue bands both above and below the horizontal axis for each of the real and imaginary parts of the coupling product $(g_R)_{b\ell}(g_L)_{s\ell}^*$. Similarly, the different sets of solutions for the $C_S^\prime$ WC translate into fig.~\ref{flavour_plots}(d) for the coupling product $(g_L)_{b\ell}(g_R)_{s\ell}^*$. These solutions do not overlap as in the case of~\ref{flavour_plots}(c), and hence we get distinct bands corresponding to the real and imaginary parts of the corresponding coupling product. As in the former cases, we provide values for these coupling products for the benchmark value $M_{V_2}=1600$ GeV:
\begin{eqnarray}
{\rm Re}((g_R)_{b\ell}(g_L)_{s\ell}^*)&\in&[-0.025,0.025],\nn \\
~{\rm Im}((g_R)_{b\ell}(g_L)_{s\ell}^*)&\in&[-0.0032,0.0032];\nn \\
{\rm Re}((g_L)_{b\ell}(g_R)_{s\ell}^*)&\in&\big([-0.0035,-0.0014], [0.0006,0.003]\big),\nn \\
{\rm Im}((g_L)_{b\ell}(g_R)_{s\ell}^*)&\in&\big([-0.025,-0.020],[0.020,0.025]\big).\nn 
\end{eqnarray}

%%%%%%%%%%%%%%%%%%%%%%%%%%%%%%%%%%%%%%%%%%%%%
\section{Collider Analysis}\label{collider}
%%%%%%%%%%%%%%%%%%%%%%%%%%%%%%%%%%%%%%%%%%%%%
In this section we study the collider prospects of $V_2$ VLQ at the LHC. We look for signals where the $V_2$ VLQ decays into a bottom quark ($b$) and a lepton ($\ell\equiv e, \mu$) with a branching ratio that depends on the corresponding coupling. We vary the coupling of $V_2$ to $b$ quark and $\ell$ from 0.1 to 0.9. As a result, the branching ratio varies from $11\%$ to $47.9\%$ for individual light leptonic channels. For further simplicity, we assume the coupling of $V_2$ to both lepton and bottom quark to be equal while that to the rest of the quarks and leptons is fixed at 0.1. Hence, the signal we consider from VLQ pair production is two $b$-jets with $P_T^{b-{\rm jet}} \ge 20$ GeV and $|\eta_{b-{\rm jet}}| \le 2.4$ and two light leptons with $P_T^{\ell} \ge 10$ GeV and $|\eta_{\ell}| \le 2.4$. The dominant backgrounds from the SM processes are $t\bar{t}$ + jets, $t\bar{t}  W^\pm$+ jets and $t\bar{t}  Z$+ jets. Furthermore, the SM process which contribute sub-dominantly are $t W^{\pm}$ + jets and $ZZ$ + jets. The SM processes like $W^+W^-$ + jets, $ZW^\pm$ + jets, $Z$+ jets and $W^\pm$+ jets contribute mildly to this analysis because we tag two $b$-jets in the final states.  We therefore do not consider these backgrounds in our present analysis. 

Both the signal and SM background processes in this analysis have been generated using {\sl Madgraph5}\;\cite{Alwall:2014hca} with the default parton distribution functions {\sl NNPDF3.0}~\cite{Ball:2014uwa}. The VLQ model file used in this analysis is obtained from {\sl FeynRules}~\cite{Alloul:2013bka}. The parton level events generated from Madgraph5 are then passed through {\sl Pythia8}~\cite{Sjostrand:2014zea} for showering and hadronization. The backgrounds and signal events are matched properly using the MLM matching scheme~\cite{Hoche:2006ph}. The detector level simulation is done using {\sl Delphes}(v3)~\cite{deFavereau:2013fsa} and the jets are constructed using fastjet~\cite{Cacciari:2011ma} with anti-$K_{T}$ jet algorithm with radius $R = 0.5$ and $P_T > 20$ GeV. The cross-section corresponding to the background processes that have been used in this analysis are provided in table~\ref{crosssection}. The signal cross-section is calculated from Madgraph at LO (leading-order).
%--------------------------------------------------------
\begin{table}[h]
\begin{center}
	\begin{tabular}{|p{4cm}|p{4cm}|}
		\hline
		Background process & cross-section (pb) \\
		\hline
		$t\bar{t}$ (NNLO + NNLL) & 815.96~\cite{Czakon:2011xx}\\
		\hline
		$t W^{\pm}$ (NLO + NNLL) & 71.7~\cite{Kidonakis:2015nna}\\
		\hline
		$t\bar{t} W^\pm (Z)$ (NLO) & 0.6448 (0.8736)~\cite{Maltoni:2015ena}\\
		\hline
		$ZZ$ (NLO) & 16.91~\cite{Cascioli:2014yka}\\
		\hline
	\end{tabular}

\caption{The cross-sections for the background processes used in this analysis are shown with the order (of QCD corrections) provided in brackets. For $t\bar{t}$, this is calculated using the Top++2.0 program up to NNLO in perturbative QCD and soft-gluon resummation up to NNLL order with the assumption that the top quark mass is 173.2 GeV.}
\label{crosssection}
\end{center}
\end{table}
%---------------------------------------------
\begin{figure}[h]
\centering
\includegraphics[keepaspectratio=true,scale=0.40]{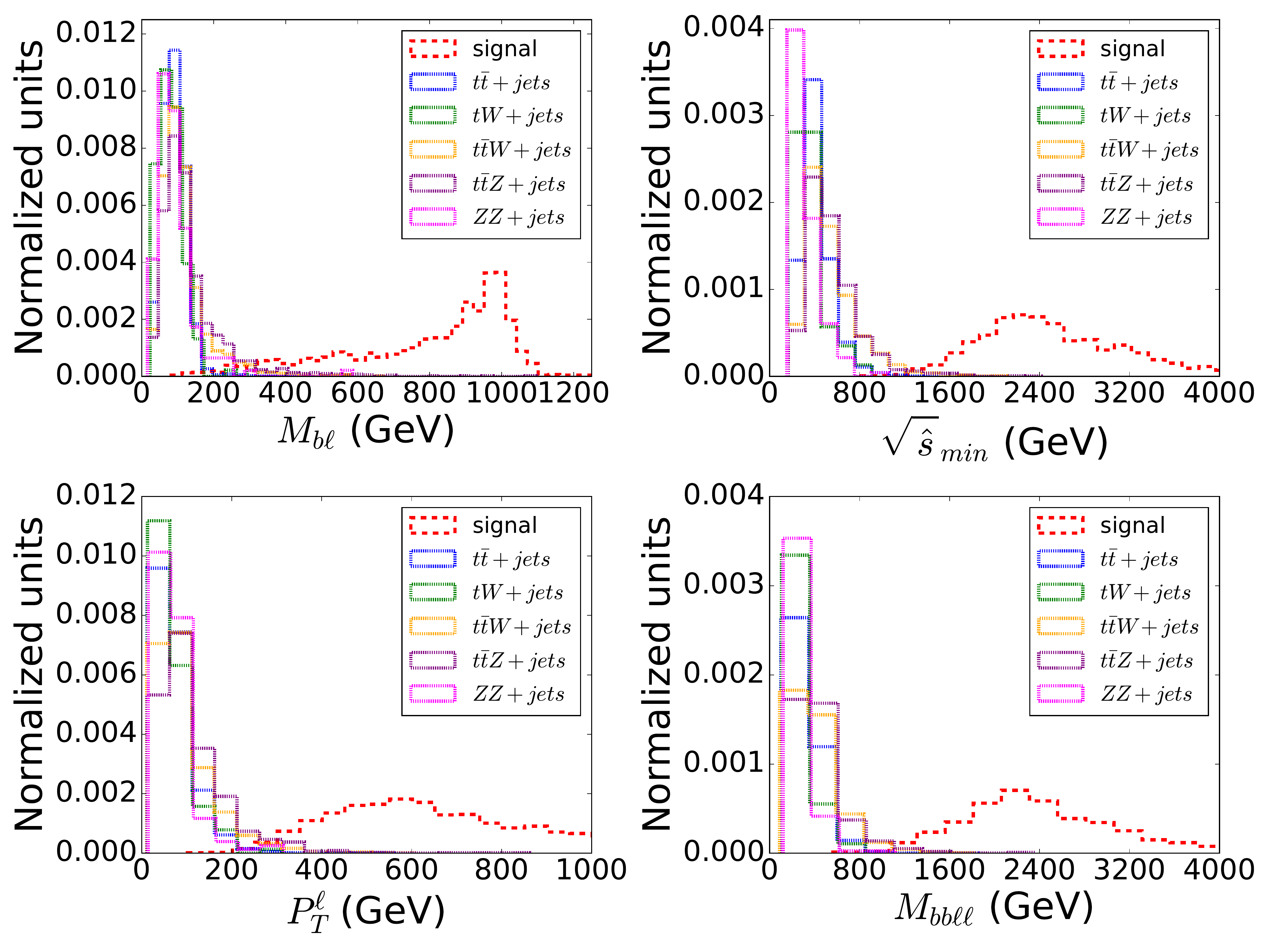}
\caption{Kinematic variables which efficiently discriminate between signal and background events are displayed here. The signal is represented by red dashed line where the $V_2$ mass is considered to be 1 TeV.}
\label{fig:obs}
\end{figure}

We have utilized some interesting kinematic variables which efficiently discriminate the signal and background events and maximizes the signal reach at the LHC. These variables are $\sqrt{\hat{s}}_{min}$~\cite{Konar:2008ei, Konar:2010ma, Swain:2014dha}, transverse momentum of the lepton, invariant mass of the $b$-jet and lepton, and the invariant mass of two $b$-jets and the di-lepton. In addition, we also make use of the di-lepton invariant mass to handle the backgrounds involving the $Z$-boson. The kinematic variable $\sqrt{\hat{s}}_{min}$ was originally proposed in order to measure the mass scale of NP produced at the LHC. It is defined as the minimum partonic CM energy that is consistent with the final state measured momenta and the missing transverse energy of the event. Mathematically, this variable is defined as,

\begin{eqnarray}\label{shatMin}
%\hspace*{-.5cm}
\sqrt{\hat{s}_{\min}(M_{inv})} = \sqrt{(E^{vis})^2 - (P_z^{vis})^2} + \sqrt{\mptvec^2 + M_{inv}^2}\;\;, 
\end{eqnarray}
where $M_{inv}$ is the sum of the masses for the ``invisible'' particles. $E^{vis} = \sum_j e^{vis}_j$ is the total energy and $P_z^{vis} = \sum_j p^z_{j}$ the total longitudinal momentum of the ``visible'' particles. In this analysis we take two $b$-jets and two leptons as our ``visible'' particles and use their momenta for calculating $\sqrt{\hat{s}}_{min}$. Since the signal we consider here does not involve any invisible particle, the missing energy in each event is very small and can solely be attributed to mis-measurement. $M_{inv}$ is also taken to be zero due to the same reason. As per our expectations, $\sqrt{\hat{s}}_{min}$ peaks at twice the mass of the LQ as shown by the red dashed distribution in fig.~\ref{fig:obs} (top panel right plot). The VLQ mass, for this representative plot, is taken to be 1 TeV. 

Similarly, the other variables like the invariant mass of the two $b$-jets and the two leptons ($M_{bb\ell\ell}$), and of one $b$-jet and corresponding lepton ($M_{b\ell}$) are also very efficient in separating the signal from the backgrounds. While the invariant mass $M_{bb\ell\ell}$ peaks at the at twice the mass of the VLQ, the variable $M_{b\ell}$ peaks at mass of the VLQ (1 TeV) as expected. Since the lepton from the VLQ is highly boosted, we also have utilized the lepton transverse momenta, $P_T^{\ell}$, as a discriminating variable. 

With the above variables we have done a cut based analysis where the following cuts are employed to maximize the signal significance,
\begin{itemize}\label{cuts}
\item $\sqrt{\hat{s}}_{min} > 1600$ GeV,
\item $P_T^{\ell} > 150$ GeV,
\item $M_{b\ell} > 150$ GeV,
\item $M_{bb\ell\ell} > 1450$ GeV,
\item $M_{\ell\ell} > 110$ GeV.
\end{itemize}
%----------------------------------------------------
\begin{figure}[h]
\centering
\includegraphics[keepaspectratio=true,scale=0.50]{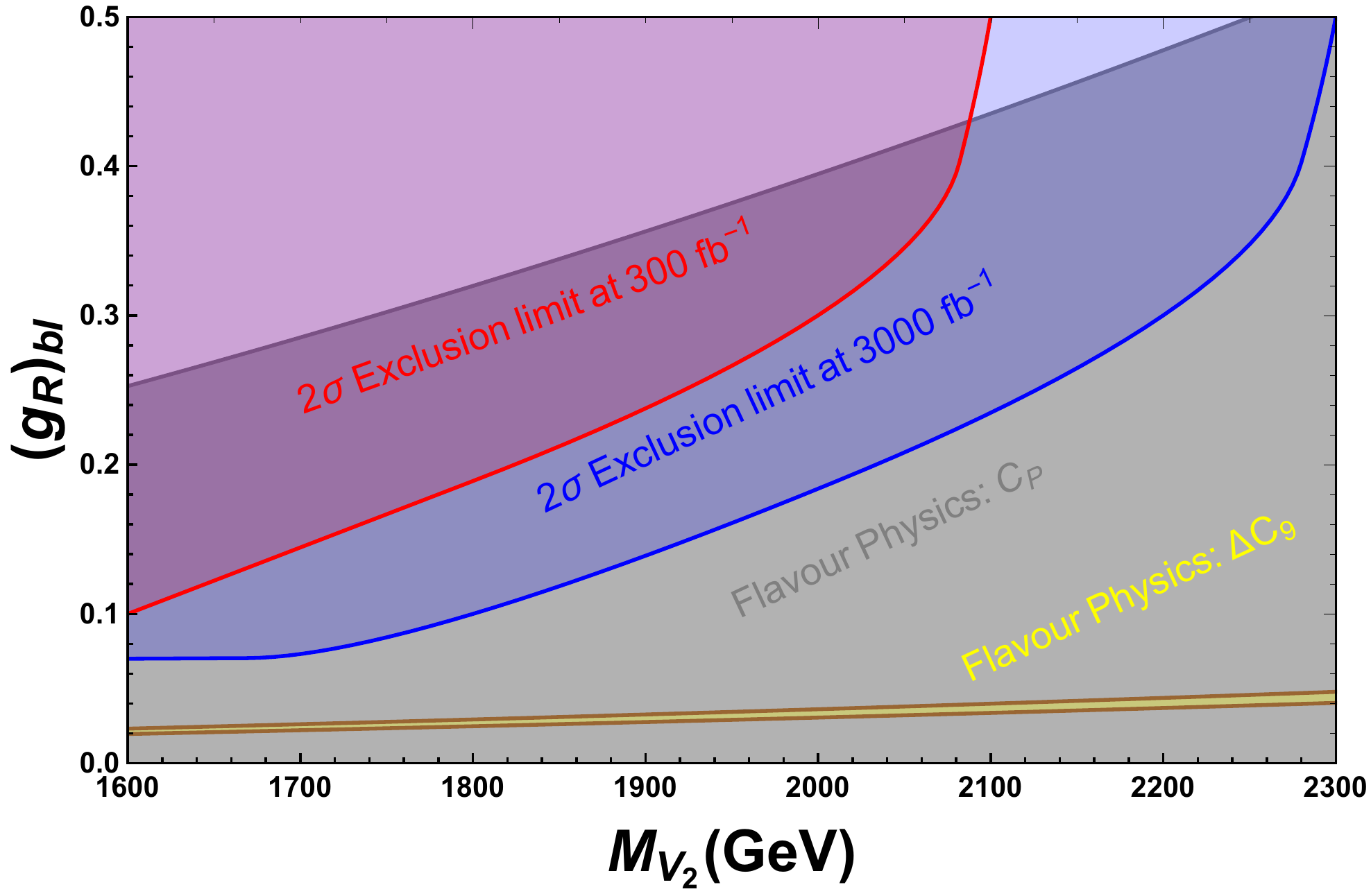}
\caption{The $2\sigma$ exclusion limits for the 13 TeV CM energy are displayed for the signal with integrated luminosities of 300 $\rm{fb}^{-1}$ (red band) and 3000 $\rm{fb}^{-1}$ (blue band) respectively. The grey band represents the constraints from the flavour physics WC $C_P$ which contribute to the $b\to s \ell^+ \ell^-$ transition as shown in eq.~\ref{WC}. The yellow band represents constraints due to the same sub-quark process coming from the $\Delta C_9$ WC.}
\label{significance}
\end{figure}
%--------------------------------------------------------------------------------
After implementing the above cuts, we have calculated the signal significance using the following formula,
\begin{eqnarray}
{\cal S} = \sqrt{2 \times [(N_S+ N_B) ~{\rm ln}(1+\frac{N_S}{N_B }) - N_S]}.
\label{eq:signi}
\end{eqnarray}
Here $N_S(N_B)$ represent the number of signal (background) events for a given luminosity after implementing the cuts mentioned above. Eq.~\ref{eq:signi} allows us to exclude the mass of the $V_2$ VLQ up to 2140 GeV for the coupling $(g_R)_{b\ell} = 0.9$ at $95\%$ C.L. for 13 TeV LHC with 300 $\rm{fb}^{-1}$ of integrated luminosity. This limit is reduced to a value as low as 1.6 TeV for $(g_R)_{b\ell} = 0.1$ (displayed in fig.~\ref{significance} with red band) at $95\%$ C.L. for 300 $\rm{fb}^{-1}$ of integrated luminosity. As is evident from the figure, the exclusion limit can go up to 2340 GeV for 3000 $\rm{fb}^{-1}$ at $95\%$ C.L. for $(g_R)_{b\ell} = 0.9$ and for $(g_R)_{b\ell} = 0.1$ the limit is 1.8 TeV which is represented by the blue band. Note that one might expect a better limit by limiting the other coupling(s) to a very small value (which, as mentioned earlier we have taken to be 0.1) so that the considered channel will get $100\%$ branching ratio. However, that limit as we have checked, is marginally better than for $(g_R)_{b\ell} = 0.9$ because even in this case also the branching ratio approaches  $100\%$. Hence, in this analysis, the limits that we have obtained for $V_2$ VLQ in mass and coupling plane from the collider study in conjunction with the flavour physics constraints are more or less optimal.

As discussed earlier in sec.~\ref{flav} using the WCs of the flavour physics observables like $R_K$, $R^{\rm low-bin}_{K^{*}}$, $R^{\rm central-bin}_{K^{*}}$ and ${\rm BR}(B_s\to\mu^+ \mu^-)$ one can obtain constraints in the VLQ mass and coupling product plane which is demonstrated in fig.~\ref{significance} by the gray region. The coupling ($(g_{R})_{b\ell}$) represented by the vertical axis is obtained by setting $(g_{R,L})_{s\ell}=0.1$ in the corresponding coupling product. It is however not possible for us to put constraints on the imaginary part of individual couplings from a combined collider and flavour point of view, since the collider analysis inherently assumes the couplings to be real. We find that part of the allowed parameter space for the real part of the coupling $(g_R)_{b\ell}$ (corresponding to a value of $(g_{R,L})_{s\ell}=0.1$) is disallowed by the collider constraints. However, the parameter space due to flavour constraints from $\Delta C_9$ (NP contribution to the WC $C_9$) is retained completely. We hence conclude that the values for $(g_R)_{b\ell}$ that fall within the yellow band in fig.~\ref{significance} represent the allowed parameter space upto $1\sigma$ with respect to the mass of the $V_2$ VLQ for all collider and flavour constraints taken together. At this point we remark in passing that, a similar analysis can also be done for $(g_L)_{b\ell}$. However, from fig.~\ref{flavour_plots} it is clear that one will not obtain common points for the real part of such a coupling after requiring $(g_{R,L})_{s\ell}=0.1$ from the flavour analysis alone (see figs.~\ref{c10pr} and~\ref{cSpr}). Moreover, most of the allowed parameter space for such a scenario will correspond to negative values of $(g_L)_{b\ell}$ and hence will have no intersection with the constraints due to the collider analysis. This will hence provide no further insight as to the allowed parameter space for such a coupling and hence we refrain from showing the corresponding plot.

%\vspace*{-.35cm}
%%%%%%%%%%%%%%%%%%%%%%%%%%%%%%%%%%%%%%%%%%%%%
\section{Conclusion}\label{con}
%%%%%%%%%%%%%%%%%%%%%%%%%%%%%%%%%%%%%%%%%%%%%
We consider a component ($V^{\frac 43}_2\equiv V_2$) of the {\boldmath${V_2}$} VLQ of electromagnetic charge $\frac 43$ which mediates $b\to s$ neutral current processes at tree level. We use the $R_K$, $R^{\rm low-bin}_{K^{*}}$, $R^{\rm central-bin}_{K^{*}}$ and ${\rm BR}(B_s\to\mu^+ \mu^-)$ data along with their $1\sigma$ errors in order to numerically solve for the involved Wilson coefficients, and, in turn, provide constraints on the product of coupling with respect to the mass of the $V_2$ VLQ. Simultaneously, we probe this VLQ at 13 TeV LHC via $b\bar{b}\ell^+\ell^-$ final state. For a reliable collider analysis we have accounted for several relevant SM background processes. Using different interesting kinematic variables and with judicious cut selections we maximise the signal significance with respect to the SM backgrounds. Our collider study reveals that it is possible to maximally exclude the mass of the $V_2$ VLQ up to 2340 GeV at 95\% C.L. at the 13 TeV LHC for an integrated luminosity of 3000 ${\rm fb}^{-1}$. In addition, our collier study reduces chunk of parameter space that is consistent with the $b\to s$ neutral current observables in the $(g_R)_{b\ell}$ coupling and VLQ mass plane for a fixed value of $(g_{R,L})_{s\ell}=0.1$.

\vspace*{0.5cm}
%%%%%%%%%%%%%%%%%%%%%%%%%%%%%%%%%%%%%%%%%%%
{\bf Acknowledgements}
%%%%%%%%%%%%%%%%%%%%%%%%%%%%%%%%%%%%%%%%%%%
AS would like to thank Ilja Dorsner for discussions regarding the implementation of the  model file for {\boldmath${V_2}$} leptoquark in {\sl FeynRules}. AKS acknowledges the support received from Department of Science and Technology, Government of India under the fellowship reference number PDF/2017/002935 (SERB NPDF).

\newpage
%\bibliographystyle{unsrt}
%\vspace*{-.45cm}
\providecommand{\href}[2]{#2}\begingroup\raggedright\endgroup

\end{document}